\documentclass{amsart}
\usepackage{graphicx}
\newtheorem{thm}{Theorem}[section]

\newtheorem{prop}[thm]{Proposition}
\theoremstyle{definition}
\newtheorem{defn}[thm]{Definition}
\theoremstyle{remark}

\numberwithin{equation}{section}

\begin{document}
\title[Dimensional analysis]{Dimensional analysis using toric ideals}
\author{M. A. Atherton}
\address{Brunel University, Middlesex, UK}
\email{Mark.Atherton@brunel.ac.uk}
\author{R. A. Bates}
\address{Rolls-Royce plc, Derby, UKJ }
\email{Ron.Bates@Rolls-Royce.com}
\author{H. P. Wynn}
\address{London School of Economics, London, UK}
\email{h.wynn@lse.ac.uk}
\begin{abstract}
Classical dimensional analysis is one of the cornerstones of
qualitative physics and is also used in the analysis of engineering
systems, for example in engineering design. The basic power product relationship in dimensional analysis is
identical to one way of defining toric ideals in algebraic geometry,
a large and growing field. This paper exploits the toric representation to provide a method for automatic
dimensional analysis for engineering systems. In particular all
``primitive", invariants for a particular problem, in a well defined
sense, can be found using such methods.
\end{abstract}
\maketitle
\section{Dimensional analysis}
Dimensional analysis has a long history. It was discussed by Newton
and provided useful intuition to Maxwell, see \cite{agost}, chapter
3. A recent paper giving a pleasant popular overview is
\cite{bolster11}. The first rigorous and most well-known treatment
is by Buckingham \cite{buck}, whose name is attached to the main
theorem. Dimensional analysis is still considered a fundamental part
of physics and is taught at an early stage in schools and colleges
as a basic part of the physics syllabus. It is often covered under a
heading of qualitative physics \cite{bhaskar}.  In engineering  it
gives a useful additional tool for the analysis of systems
\cite{gibbings}. It is used in engineering design
and in the formal design of engineering experiments \cite{gibbings1} \cite{grove}. It has also been used in
economics \cite{barnett}. For an interesting recent application
to turbulence and criticality see \cite{chapman09}
\cite{chapman091}.

We shall give an algebraic development of dimensional analysis based on the theory of toric
ideals and toric varieties. Although this is essentially a
reformulation, the algebraic theory itself is by no means elementary.
The theory of toric ideals is a live branch of algebraic geometry.
We have used \cite{sturm} and the recent comprehensive volume
\cite{cox}. We shall see that the methods give all ``primitive"
invariants for a particular problem, in a well-defined sense.

Within mathematical physics dimensional analysis can also be seen as
an elementary application of the theory of Lie groups and
invariants, when the group is the scale group defined by
multiplication. We shall draw on \cite{olver} in the penultimate
section.

The basic idea of dimensional analysis is that physical systems use
fundamental quantities, or units, of mass (M), length (L) and time
(T). To this list may been added various others such as temperature
(K) and current (I), depending on the domain. The extent to which
new fundamental quantities can be expressed in terms of $M,L,T$ goes to
the heart of physics but we shall not delve deeply.
Mathematical models for physical systems use so-called {\em derived}
quantities such as: force, energy, momentum, capacity etc.
Dimensional analysis tells us that each one of these quantities has
units which have a power product representation. Table 1 gives a few
examples from  mechanics.

\vspace{3mm}
\begin{center}
\begin{tabular}{|l|l|}
  \hline
  Quantity & units \\
  \hline
  momentum & $MLT^-1$   \\
  force &  $MLT^{-2}$\\
  work & $ML^2 T^{-2}$  \\
  energy & $ML^2T^{-2}$ \\
  pressure & $ML^{-1}T^{-2}$ \\
  density & $ML^{-3}$ \\
  volumetric flow & $L^3 T^{-1}$    \\
  \hline
\end{tabular}
\vspace{3mm}

Table 1. Some basic derived quantities

\end{center}

\vspace{3mm}
\noindent We note that the formulae for the expression of derived units have
integer powers. This is critical for our development: it makes them
{\em algebraic} in the sense of polynomial algebra.

In a physical system we may be interested in a special collection of
derived quantities. The task of dimensional analysis is to derive
dimensionless variables with a view to finding, by additional theory
or experiment, or by both, the relationship between these
dimensionless quantities. As mentioned, the key theorem in the area
is due to Buckingham. In this section we explain it with an
example, leaving a  more detailed discussion until later.

Rather than use the $M,L,T \ldots $ notation we assume that there
are some basic quantities of interest which we label $z_1, z_2,
\ldots$. Each quantity is assumed to have the scaling property,
namely if the fundamental units, which we now call $t_1, t_2,
\ldots$ are scaled up or down this induces a transformation
on the $z_i$. Whether this means simply a change in units or actual
physical scaling of the system is sometimes unclear
in the literature, but we shall prefer the latter interpretation.

As  example, if $z_1$ is force and the fundamental units are mass
$(t_1)$, length $(t_2)$ and time $(t_3)$, then the scaling
transformation is
$$z_1 \rightarrow t_1 t_2 t_3^{-2} z_1.$$
With a collection of derived quantities we have one such transformation
for each $z_j$. A slightly more realistic formulation is to introduce
non-zero constants $c_j$ so that in this case we would have
$$z_1 \rightarrow c_1t_1 t_2 t_3^{-2} z_1,$$
but this would make little difference to our derivations

Here is a well-known example which we shall use as our running example. It concerns a body in a
fluid and the quantities of interest are fluid density ($z_1$),
fluid velocity ($z_2$), object diameter ($z_3$), fluid viscosity
($z_4$)and fluid resistance ($z_5$). Taking the units into account
the transformation is:
\begin{equation}
\left(\begin{array}{c}
  z_1 \\
  z_2 \\
  z_3 \\
  z_4 \\
  z_5
\end{array}
\right) \rightarrow \left(
  \begin{array}{c}
    t_1t_3^{-3} z_1 \\
    t_2^{-1}t_1 z_2 \\
    t_3z_3 \\
    t_1t_2^{-1}t_3^{-1}z_4 \\
    t_1t_2^{-2}t_3 z_5 \\
  \end{array}
\right) \label{trans}
\end{equation}
After a little algebra, or formal use of Buckingham's theorem, we can derive dimensionaless quantities
$$y_1 = z_1z_2z_3z_4^{-1}, \;\; y_2 = z_1^{-1}z_2^{-2}z_3^{-2}z_5.$$
The first quantity is Reynolds number. The term {\em dimensionless} is
interpreted by saying that replacing each $z_j$ by the $y_j$
in the  transformation $\rightarrow$ in \ref{trans}, leaves the
expression unchanged: the $y_j$ are rational invariants of the
transformation. The dimensionless principal, for our example, embodied in the
Buckingham theorem is that {\em any} function $F$ of $x_1, \ldots,
x_5$ which is invariant under $\rightarrow$ is a function of $y_1$
and $y_2$ which we write: $F(y_1,y_2)$.

We now sketch the traditional method. The transformation $\rightarrow$ can be coded up by capturing the
exponents in the power products. This gives
$$A =
\left(
  \begin{array}{rrrrr}
    1 & 0 & 0 & 1 & 1 \\
    0 & -1 & 0 & -1 & -2 \\
    -3 & 1 & 1 & -1 & 1 \\
  \end{array}
\right). \label{K}
$$
This matrix has rank 3 and we can find a full rank $2 \times 5$
kernel matrix $K$. Namely, a  $K$ which has rank 2 such that $A^TK=0$. This is readily computed using
existing functions in computer algebra such as the ``nullspace" command on Maple. We obtained
$$K =
\left(
  \begin{array}{rrrrr}
    1 & 1 & 1 & -1 & 0 \\
    -1 & -2 & -2 & -2 & 1 \\
  \end{array}
\right). \label{gens}
$$
The key point is that the rows of this $K$ give the exponents of $z_1, \ldots, z_5$ in
$y_1$ and $y_2$. However,  we can also derive alternative $K$. For example,
$$
K' = \left(
  \begin{array}{rrrrr}
    1 & 1 & 1 & -1 & 1 \\
    0 & -1 & -1 & -1 & 1 \\
  \end{array}
\right).
$$
This gives an alternative to $y_2$, above, namely $y_3 =
t_2^{-1}t_3^{-1}t_4^{-1}t_5$. The toric approach clarifies, among other issues, the immediate
problem of the choice of $K$ which this example exposes.

\section{Toric ideals}
\label{toric} Algebraic geometry is concerned with {\em ideals} and
their counterpart algebraic {\em varieties}. We give a very short
description here. (Note that we shall use $x$ for variables in an abstract algebraic setting
reserving $z$ for ``real" problems.) A standard reference is \cite{cox}. We start with the
ring of all polynomials in $n$ variables $\{x_1, \ldots, x_n\}$ over
a field $k$: $k[x_1, \ldots, x_n]$. A set $I$ of polynomials is an
ideal if $F \in I$ implies $s(x)f(x)$ is in $I$ for any $s(x)$ in
$k[x_1, \ldots, x_n]$. By a theorem of Hilbert all ideals are
finitely generated. That is we can find a set of polynomials
$f_1(x), \ldots f_m(x)$ such that any $f(x)$ in $k[x_1, \ldots,
x_n]$ can be written $f(x) = s_1(x)f_1(x) + \cdots + s_m(x)f_m(x)$
for some $\{s_i(x)\}$ in $k[x_1, \ldots, x_n]$. An ideal $I$ gives a
variety as the set of $x$ such that $f(x) =0$ for all $f(x) \in
I$. The other identity that $I$ is the set of all polynomials zero
on the variety is not always true, but for the purposes of this
paper we will alternate freely between varieties and ideal. It will
also be enough to work within the field $Q$ of rationals.

Modern computational algebra has benefitted hugely from the theory
of Gr\"obner bases and the algorithms that grew out of the theory,
notably the Buchburger algorithm. We will need one more concept, that
of a monomial term ordering, or term ordering for short. Monomials $x^{\alpha} = x_1^{\alpha_1}
... x_n^{\alpha_n}$, where $\alpha = {\alpha_1, \dots, \alpha_n} \geq 0$ ie $\alpha_i \geq 0,\;
i=1,\ldots,n$, drive the theory. A monomial term
ordering, written  $x^{\alpha}\prec x^{\beta} $ between is a total
(linear) ordering with the addition condition: $x^{\alpha}\prec
x^{\beta}$ implies $x^{\alpha+\gamma}\prec x^{\beta+ \gamma}$, for
all $\gamma \geq 0$. Since such an ordering is linear every
polynomial $f$ has a leading term $LT_{\prec}(f)$. If we fix the monomial
ordering, $\prec$, the Gr\"obner basis $G_{\prec}=\{g_1(x), \ldots,
g_m(x)\}$ of an ideal $I$ with respect to $\prec$ is a basis such
that the ideal generated by {\em all} monomials in the ideal is the
same as that generate by the leading terms of $G_{\prec}$.
Given $I$ and $\prec$ the Buchburger algorithm delivers $G_{\prec}$.
We will be concerned with the set of all Gr\"obner bases as $\prec$
ranges over all monomial term orderings. This is called the {\em
fan} and is finite, although is can be very large.

One of the main definitions of a toric ideal fits perfectly with the
power product transformations of dimensional analysis. It is this
observation which motivates this paper. We will emphasize the
connection by using the same notation: $\{t, y, A\}$, with $x$ or $z$ according
to emphasis, but with $t$, $y$ and $A$ used in both the pure algebra and physical
theories.

The following development can be taken from an number of books, but \cite{sturm} is our main
source. The main steps in the definition are.

\begin{enumerate}
\item The polynomial ring over $n$ variables $k[{\bf x}]=k[x_1, \ldots, x_n]$.
\item A $d \times n $ matrix $A$ with columns labeled $\bf{a}_1,
\dots, \bf{a}_d$.
\item Variables $t_1,\ldots, t_d$  and the Laurent ring generated by
the $t_i$ and the inverses $t_i^{-1}$. We write this as
$$k[{\bf t},{\bf t}^{-1}]=k[t_1,\ldots, t_d, t_1^{-1}, \ldots t_d^{-1}].$$
\item A power product mapping from $k[{\bf x}]$ to $k[{\bf t},
{\bf t}^{-1}]$ defined by $A$:
$$x_i \rightarrow t^{{\bf a}_i}$$
\end{enumerate}

The kernel of the mapping in (4) above is the toric ideal. It can be
considered as the ideal obtained by formally eliminating ${\bf t}$ t from the
ideal:
$$\langle x_i -t^{{\bf a}_i}, i=1, \ldots, n \rangle$$

The following paragraph should be considered as a theorem.

The generators of the toric ideal ideal are related to the kernel of $A$ in the
follow way. The generators are all so-called {\em binomials}
$${\bf x}^{\bf{u}} - \bf{x}^{\bf{v}},$$
where $\bf{u}$ and $\bf{v}$ are non-negative integer vectors with
the property that $$A {\bf u} = A{\bf v}. $$

The last equation can be written $A(\bf{u}-\bf{v})=0$, which is
equivalent to $\bf{u}-\bf{v}$ being in the kernel of $A$.

The connection  with dimensional analysis should now be clear.
Let us put dimensional analysis on a similar notational footing, only using $z$ instead of $x$.
Start with a $d \times n $ matrix $A$
with columns $\{ {\bf a}_i \}$. The general form of the mapping $\rightarrow$ in \ref{trans} becomes
\begin{equation}
z_i \rightarrow  {\bf t}^{{\bf a}_i} z_i, \; i=1,\ldots, n
\end{equation}
We can write this in matrix terms as
\begin{equation}
\bf{z} \rightarrow \bf{t}^A z \label{arrow}
\end{equation}
Now, suppose we have a possible invariant $y_j$.
Using  ${\bf u},
{\bf v}$ to denote integer vectors with non-negative entries to distinguish the positive from the
negative exponents and write
$$y_j = \bf{z}^{{\bf u}_j}\bf{z}^{-{\bf v}_j} $$

The condition to be an invariant is that substituting each $z_j$ by $y_j$
in the right hand side of  \ref{arrow} for $z$ leaves $y_j$
unchanged. But the condition for this is
$${\bf z}^{{\bf u}}{\bf z}^{-{\bf v}} = ({\bf t}^A{\bf z})^{{\bf u}}({\bf t}^A{\bf z})^{-{\bf v}},\;\ j=1,\ldots, d,$$
which is equivalent to
$$A{\bf u}_j - A {\bf v}_j =0,\;\ j=1,\ldots, d,$$
exactly the toric condition.
We have proved our main result:
\begin{thm}
A variable $y$ is a dimensional invariant in a system
defined by a  matrix $A$, with derived variable $\bf{z}$, if any
only if it takes the form
$$y = \bf{z}^{\bf{u}} \bf{z}^{-\bf{v}}$$
where $u$ and $v$ are non negative integer vectors such that
$A{\bf u} = A {\bf v}.$ Moreover the set of all quantities
$${\bf z}^{{\bf u}} - {\bf z}^{{\bf v}},$$
 is the toric ideal $I_A$ with generator matrix $A$.
\end{thm}

A brief summary is to say that the set of all dimensional quantities
$y$ associated with $A$ are exactly those given by the toric ideal $I_A$.

We can give
a minimal set of generators for the toric ideal of our running
example. We use the ``Toric" function on the computer algebra package
CoCoa \cite{CocoaSystem}, which takes the matrix $A$ as input. Simply to ease the notation in the use of computer
algebra we use
$a, \ldots, e$, for $z_1, \ldots,z_5$. The script with
output is.

\vspace{3mm}

  Use $R ::= QQ[a,b,c,d,e]$;

  Toric([[1,0,0,1,1],[-3,1,1,-1,1],[0,-1,0,-1,-2]]);

Ideal$(-d^2 + ae, abc - d, bcd - e)$

\vspace{3mm}
By the theorem, given any generator we have a invariant. Thus $-d^2 + ae$
yields $\frac{ae}{d^2}$. Thus we have gives three invariants:
$$\frac{ae}{d^2},\; \frac{abc}{d},\;\frac{bcd}{e}$$
We see that the second two ideal generators give exactly the
dimensional variables from the kernel matrix $K'$, above.
A key point is that the toric ideal may have more generators than
the rank of the kernel in Buckingham's theorem. The next section
explains why this is so.

\subsection{Saturation and Gr\"obner bases}
\label{sat}
To summarise, the toric version of dimensional analysis says that we
can generate  dimensionless quantities from the toric ideal which is
the elimination ideal of the original power product representation,
being careful to use elimination in the proper algebraic sense.

A lattice ideal associated with an integer defining matrix $A$ is
the ideal based on a full rank kernel matrix.  That is if  $A$ is $d
\times n$ with rank $d$ then we find an integer $n \times d-n$
matrix $K$, with rank $n-d$ with rows ${\bf k}_1, \ldots,
{\bf k}_{n-d}$ with $A^TK=0$.

The corresponding lattice ideal is
generated by $\{{\bf t}^{{\bf k}_j}\}$. For our first $K$ in \ref{K}
{\em lattice ideal} has two generators:
$$\langle z_1z_2z_3z_4^{-1}, z_1^{-1}z_2^{-2}z_3^{-2}z_4^{-2}z_5 \rangle.$$

But, as we have seen, this has one fewer generators than
the toric ideal. However, given any such lattice ideal we can obtain
the toric ideal using a process called {\em saturation}. The process
has two steps. Fix the defining matrix $A$ and let $I_A$ be a
lattice ideal associated with $A$.
\begin{enumerate}
\item Select a dummy variable $s$ and adjoin to the lattice ideal
the generator $s\prod_{j=1}^n x_j+1$. That is form the union
$$I_K^* = I_K \cup \langle s\prod_{j=1}^n x_j+1 \rangle.$$
\item Eliminate $s$ from $I_K^*$ to give the toric ideal for
$\{x_1, \ldots, x_n\}$. That is, the toric ideal is obtained as the
elimination ideal for $\{x_1, \ldots, x_n\}$.
\end{enumerate}
The process of elimination in this saturation process is a formal procedure
and leads to a {\em reduced Gr\"obner basis} of the toric ideal which in general depends in general on
the monomial ordering used in the elimination algorithm.

This process gives an explanation for the fact that the toric ideal
contains, but is not necessarily equal to the lattice ideal.
Recall that unions of ideals is mirrored by intersections of
varieties. The addition condition $s\prod_{j=1}^n x_j+1=0$ giving
the variety defined by $I_A^*$ forces all the $x_j$ to be
nonzero. This property is inherited by the toric ideal. It
implies that if any $x_j =0$ then all $x_j$ and zero. That is to say,
saturation removes the principal axes and all axial subspaces.

This gives a nice physical interpretation. If we exclude the origin, then
for the toric variety associated with the toric ideal {\em must not
contain any other zeros}. Translated into the original $z_j$
variables, the toric ideal description of
the dimensionless quantities is appropriate when {\em non of the
defining variables $z_j$ is allowed to be zero}. This removal of zeros
is intimately connected with the abstract definitions of toric
varieties based on the concept of a torus in complex variables, but we do
not develop this here, see \cite{cox}.

\subsection{The Gr\"obner fan, primitive invariants and the Graver basis}
\label{prim}
A natural question given the ease of computing invariants using
toric methods is whether the invariants obtained in this way are in
some sense minimal. This turns out to be the case. We can illustrate
this with our example. A little inspection of the basis $-d^2 + ae,
abc - d, bcd - e$ shows that we cannot get simpler invariants from
this basis by multiplication (or division) : eg if
$$y_1 = \frac{ae}{d^2},\; y_3= \frac{bcd}{e}$$ then $y_1y_2 = \frac{abc}{d}$
which, although a new invariant, is not obtained by reducing the
numerator or denominator of any of the original invariants.
\begin{defn}
A basis element $z^u-z^{v}$ of $I_A$ is called is called {\em primitive} if there is no basis element
invariant $z^{u'}-z^{v'}$ such that such that $z^{u'}$ divides $z^u$ and
$z^{v'}$ divides $z^v$. We call an invariant $y = z^u z^{-v}$ {\em primitive} if and only if $z^u-z^{v}$
is primitive as a basis element of $I_A$.
\end{defn}
Lemma 4.6 of \cite{sturm} is
\begin{thm}\label{primitive}
Every invariant obtained from as a reduced Gr\"obner basis of $I_A$
is primitive.
\end{thm}
Note that in what follows we are  a little lazy in not disassociating an invariant from its inverse.

As mentioned, as we range over all monomial term orderings defining
the individual Gr\"obner basis we obtain the complete Gr\"obner
fan and by the lemma and our definition all resulting primitive invariants are primitive.
This union of bases is called the {\em universal} Gr\"obner basis and the computer
programme Gfan is recommended to compute the fan \cite{gfan}.

We return to our running example. If we put the G-basis
element $\langle-d^2+ae, abc-d, bcd-e \rangle$ into Gfan we obtain
the full fan as
$$\langle bcd -e,ae-d^2, abc-d \rangle,\;
\langle e-bcd, abc-d \rangle,\; \langle d^2 - ae, bcd - d, abc-d
\rangle$$
$$\langle d-abc, ab^2c^2-e \rangle,\; \langle e-ab^2c^2, d-abc \rangle,$$
the first of which is the input basis.

The universal Gr\"obner basis of distinct basis terms (ignoring the sign change) copy is:
$$bcd-e, ae-d^2,abc-d, ab^2c^2-e$$ and we have a new primitive invariant: $\frac{ab^2c^2}{e}$.

The set of {\em all} primitive polynomials, which may be larger than that
giving the union of the basis elements in the fan, is called the {\em
Graver} basis. Algorithm 7.2 of \cite{sturm} can be used for this.

Briefly, the method starts by constructing from $A$ an extended matrix
called the Lawrence lifting:
$$
\tilde{A} = \left[
\begin{array}{cc}
  A & 0 \\
  I & I
\end{array}, \right]
$$
where the zero is a $d \times n $ zero matrix
and $I$ is a $d \times d$ identity matrix. Then
introducing $n$ more derived variables to make a set $z_1, \ldots, z_n, z_{n+1}, \ldots z_{2n}$ a toric
ideal is constructed using $\tilde{A}$. Finally, set $z_{n+1} = \cdots = z_{2n} =1$

The method is conveniently set out in the help screen of
``ToricIdealBasis" on Maple. After inputting
$$ A =
\left[\begin{array}{rrrrrrrrrr}
  1 & 0 & 0 & 1 & 1 & 0 & 0 & 0 & 0 & 0 \\
  0 & -1 & 0 & -1 & -2 & 0 & 0 & 0 & 0 & 0 \\
  -3 & 1 & 1 & -1 & 1 & 0 & 0 & 0 & 0 & 0 \\
  1& 0 & 0 & 0 & 0 & 1 & 0 & 0 & 0 & 0 \\
  0 & 1 & 0 & 0 & 0 & 0 & 1 & 0 & 0 & 0 \\
  0 & 0 & 1 & 0 & 0 & 0 & 0 & 1 & 0 & 0 \\
  0 & 0 & 0 & 1 & 0 & 0 & 0 & 0 & 1 & 0 \\
  0 & 0 & 0 & 0 & 1 & 0 & 0 & 0 & 0 & 1
\end{array} \right],
$$
we use the commands
\vspace{3mm}

$ zs := [seq(z[i],i=1..10)];$

$T:=ToricIdealBasis(A, zs, plex(op(zs)), method = 'hs', grading=grd);
$

$
G:= subs([seq(zs[i]=1,i=6..10)],T);
$
\vspace{3mm}

\noindent This yields
$$ \langle z_2z_3z_4-z_5, z_1z_5 - z_4^2, -z_4+z_1z_2z_3,-z_5+z_1z_2^2z_3^2 \rangle,$$
In this case the set is the same as given by the fan. That is, the
universsal Gr\"obner basis and the Graver basis are the same.

\section{Further examples}
\label{applications}

For each of the examples we give the derived quantities using the
classical notation, (i) the $A$ matrix (ii) a single toric ideal
basis give by the default function on CoCoa and (iii) a full set of
primitive basis elements, that is the Graver basis, given by the
maple `ToricIdealBasis" command. From this a full set of primitive
invariants is immediate. It turns out that for all except one of our
examples (windmill) the Graver basis is also the universal Gr\"obner
basis. We try to mention when we find well-known invariants.

\subsection{Windmill}\label{wind}
This standard problem is taken from \cite{gibbings}, Section 9.3.
(We have changed $d$ there to $D$).

It concerns a simple windmill widely used to pump water. The table
is \vspace{3mm}
\begin{center}
\begin{tabular}{|l|l|}
  \hline
  shaft power, $P$ & $ML^2Y^{-2}$ \\
  diameter, $D$ & $L$ \\
  wind speed, $V$ & $LT^{-1}$ \\
  rotational speed, $n$ & $T^{-1}$ \\
  air density $\rho$ & $ML^{-3}$ \\
  \hline
\end{tabular}
\end{center}
\vspace{3mm}
The $A$-matrix is
$$\left(
    \begin{array}{rrrrr}
      1 & 0 & 0 & 0 & 1 \\
      2 & 1 & 1 & 0 & -3 \\
      -3 & 0 & -1 & -1 & 0 \\
    \end{array}
  \right)
$$
In the $a,b \ldots $ notation we obtain, from Cocoa, a basis with 4
terms :
$$\langle bd - c, b^2c^3e - a, c^5e - ad^2, bc^4e - ad \rangle$$
The first entry give a dimensionless quantities discussed in the book:
$$\frac{V}{nD}.$$

The universal Gr\"obner basis obtained from the Gfan gives five
terms
$$\langle bd - c, b^2c^3e - a, c^5e - ad^2, bc^4e - ad, b^5d^3e-a \rangle$$
The last of these is also discussed in the book; it gives the
invariant
$$\frac{P}{\rho n^3 d^5}.$$

A full set of 7 primitive invariants, the Graver basis, is
$$\langle bd - c, b^2c^3e - a, c^5e - ad^2, bc^4e - ad, b^5d^3e-a,b^4cd^2e-a,bc^4e-ad \rangle,$$
Since $\mbox{rank}(A)=3$ there are only two algebraically
independent invariants. The standard argument may suggest testing
the relationship between any two independent invariants, for example
in a wind tunnel. An important question, which should be the subject
of further research, is say which two or, more generally,  whether
the dimensional analysis is sufficiently trusted to test only one
pair and infer other relationships from the algebra.

\subsection{Forced convection}
The interest is in the following derived quantities: the forced
convection coefficient $h$, the velocity, $u$, the characteristic
length of the heat transfer surface $L$, the conductivity of the
fluid $k$, the viscosity, $\mu$, the fluid specific heat capacity,
$c$ and the fluid density, $\rho$. The fundamental dimensions are
$M,L,T$ and two new ones temperature ($K$) and energy ($J$). With
columns in the order of the listed the rows in the units order the
$A$-matrix is
\vspace{3mm}
$$
\left(
  \begin{array}{rrrrrrr}
    0 & 0 & 0 & 0 & 1 & -1 & 1 \\
    -2 & 1 & 1 & -1& -1 & 0 & -3 \\
    -1 & -1 & 0 & -1 & -1 & 0 & 0 \\
    -1 & 0 & 0 & -1 & 0 & -1 & 0 \\
    1 & 0 & 0 & 1 & 0 & 1 & 0 \\
  \end{array}
\right).
$$
\vspace{3mm}
Resorting to the $a,b, \ldots$ notation we have from, Cocoa, the ideal
$$\langle ac - d, ef - d, bcg - e, bfg - a, ae - bdg \rangle,$$ giving
invariants:
$$\frac{ac}{d}, \; \frac{ef}{d},\; \frac{bcg}{e},\;\frac{bfg}{a},\;\frac{ae}{bdg}$$
The first three of these are well-known invariants:
\begin{eqnarray}
\mbox{Reynolds number}:  R & = & \frac{\rho u L}{\mu} \\
\mbox{Nusselt number}, N & = &\frac{h L}{k} \\
\mbox{Prandtl number}, P & = & \frac{\mu c}{k}
\end{eqnarray}
In preparing this paper it was pleasing to obtain these directly from the computer on the first run.
The full set of 7 primitive basis elements is
$$\langle ac - d, ef - d, bcg - e, bfg - a, ae - bdg, bcfg-d,ac-ef \rangle$$
The simplest of the ``new" primitive invariants is from $ac-ef$:
$$\frac{hL}{\mu c},$$
which is the Reynolds/Nusselt.

\subsection{Electrodynamics}
\label{electro} As an exercise we take six basic quantities for
electro-dynamics, used the literature to give some expression in
terms of mass $(M)$, length $(L)$, Time $(L)$ and current ($A$). We
do not have any particular elctromagnetic device in mind, but simply try to find
some dimensionless quantities. The table below gives one version:
\vspace{3mm}
\begin{center}
\begin{tabular}{|l|l|}
  \hline
  Quantity & units \\
  \hline
  charge &  $TA$\\
  potential & $ML^2T^{-3}A^{-1}$   \\
  capacitance & $M^{-1}L^{-2} T^4 A^2$  \\
  inductance & $ML^2T^{-2}A^{-2}$ \\
  resistance & $ML^2T^{-3}A^{-2}$ \\
  \hline
\end{tabular}
\end{center}
\vspace{3mm} The $A$-matrix is
$$A =
\left(
  \begin{array}{rrrrr}
    0 & 1 & -1 & 1 & 1  \\
    0 & 2 & -2 & 2 &2  \\
    1 & -3 & 4 & -2 & -3  \\
    1 & -1 & 2 & -2 & -2  \\
  \end{array}
\right)
.
$$
CoCoa gives
$$\langle bc - a, -ce^2 + d, -ae^2 + bd \rangle .$$
Note that $A$ only has rank 3. It turns out that this is a complete
list of primitive basis elements.

\subsection{Quantum}
Toric ideals are embedded in advanced models in physics but one can
get some way with simple dimensional analysis. This example is given
in some form by a number of authors. We found \cite{smith}, section
1.3.1, useful. The hydrogen atom consists of a proton and a
neutron and the Bohr radius is the distance between them. We have
used slightly non-standard notation. In a somewhat cavalier manner
we have introduced the speed of light as derived quantity.

\vspace{3mm}
\begin{center}
\begin{tabular}{|l|l|}
\hline
mass of electron, $m_e$ & $M$ \\
Bohr radius,  $a_0$ & $L$  \\
energy, $E$ & $ML^{-2}T^{-2}$  \\
Plank's constant, $\hbar$ & $ML^2T^{-1}$ \\
permitivity of vacuum (squared), $e^2$ & $ML^3T^{-2}$ \\
speed of light, $c$ & $LT^{-1}$ \\
\hline
\end{tabular}
\end{center}
\vspace{3mm}

Then the $A$-matrix is

$$
\left(\begin{array}{rrrrrr}
    1 & 0 & 1 & 1 & 1 & 0 \\
    0 & 1 & 2 & 2 & 3 & 1 \\
    0 & 0 & -2 & -1 & -2 & -1\\
  \end{array}
\right)
$$
The ideal is
$$ \langle -df+e, bc - df, abe - d^2, -cd^2+ae^2, abf - d, af^2 - c, aef - cd \rangle, $$
(the algebraic $e$ is $e^2$ and the algebraic $c$ should not be
confused with the speed of light). The first terms gives an
invariant called the ``fine structure constant"
$$ \frac{e^2}{\hbar c}.$$
If we take the third term and interpret $\frac{x^2}{uvy}$ being
invariant as stating that $v = \mbox{constant}\times\frac{x^2}{uy}$, then we have
a well known formula for $a_0$ interpreted as the size of the
hydrogen atom:
$$a_0 = \mbox{constant} \times \frac{\hbar^2}{m_e e^2}.$$
We cannot resist stating that the sixth basis element, $af^2-c$
gives
$$E = \mbox{constant} \times m_ec^2.$$

The Graver basis gives a full set of 10 primitive invariants for the
hydrogen atom is
$$\langle -df+e, bc - df, abe - d^2, -cd^2+ae^2, abf - d, af^2 - c, aef - cd, bc-e, abf^2-e,-f^2+ab^2c
\rangle $$ It is not known whether this list has been given
explicitly before.

\section{Group invariance} Dimensional analysis should be considered
as a special case of the theory of groups invariance and in an
attempt to suggest a natural generalisation we very briefly sketch
the theory of invariants.

We start with the action of a Lie group $G$ acting on a manifold $M$
in $R^d$. The manifold will be our model and the group
something to do with our physical understanding of the physics being
modelled. The {\em orbit} of ${\mathcal O}(x)$ is a point $x$ in $M$
be the set of all $g(x)$ for all $g$ in $G$. If $M$ is  invariant
under $G$ then ${\mathcal O}(x) \subset M$. This sets up an
equivalence relations with members of $M$ in the same orbit being
equivalent. The collection of equivalence classes is denoted by the
quotient $M/G$ and the {\em projection} $\pi: M \rightarrow M/G$
maps every member of of $M$ into its correct equivalence class. If
we are lucky then $M/G$ is a manifold in its own right and we say
that $G$ acts {\em regularly} on $M$. Also, the mapping $\pi$ can be
used to set up a coordinate system on $M/G$ and note that $\pi$
itself is an invariant. This discussion leads naturally to the
following
\begin{prop}
Let a group $G$ act regularly on a manifold $M$. Consider a manifold
defined by a smooth function $F$ is a set $S_F = \{x|f(x)=0$. Its is
$G$-invariant is and only there is a function $F^*$ defining a
smooth sub-manifold $S_{F^*} = \{y|F^*(y) = 0\}$ on $M/G$ such that
$$S_{F^*} = \pi(S_F),$$ where $\pi$ is the projection form $M$ to
$M/G$.
\end{prop}

A one parameter Lie group $G$ shifts a point $x$ along an integral
curve $\Psi(\epsilon,x)$ called a {\em flow} If we expand
$\Psi(\epsilon, x)$ in a Taylor expansion in $\epsilon$ we obtain:
$$\Psi(\epsilon) = x + \epsilon \xi(x) + O(\epsilon^2).$$
The term $\xi(x) = (\xi_1(x), \ldots, \xi_d(x)$ defines a vector
field and we can write $v$ in local coordinates in classical
$$v = \xi_1(x) \frac{\partial}{\partial x_1} + \cdots + \xi_1(x) \frac{\partial}{\partial
x_d}$$

A function $\psi$ is an invariant if $v\psi=0$ or
$$\xi_1(x) \frac{\partial \psi }{\partial x_1} + \cdots + \xi_1(x) \frac{\partial \psi}{\partial
x_d}=0.$$ This is a first order partial differential equation which can
be solved by writing down
$$\frac{dx_1}{\xi_1(x)} = \cdots = \frac{dx_1}{\xi_d(x)},$$
namely by the methods of characteristics. The solutions take the
form:
$$\psi_1(x) = c_1, \cdots , \psi_m(x) = c_m,$$
where the $\psi_j$ are the invariants.

In our notation $\epsilon$ becomes $t$ and the mapping $\rightarrow$
in (\ref{trans}) is
$$\Psi({\bf t},{\bf z}) = {\bf t}^A {\bf z}.$$
Matrix partial differentiation with respect to ${\bf t}$, and setting
all $t_i = 1$ gives the infinitesmal generators:
$${\bf v} = A \frac{\partial}{\partial {\bf z}}.$$
An interpretation of the toric variety is as characterising the
orbits of the group, as discussed above. We have not formally proved the Buckingham
theorem,  but drawing on the above discussion it is give as Theorem 2.22 in \cite{olver}.

\section{Discussion}
We have seen that the toric ideal method, via the Graver basis, is a
fast way to compute all primitive invariants in dimensional
analysis. There are three areas of further study which this
suggests.

The first area arises from the possibility that different physical
systems may yield different types of toric ideal or variety. The most
important general class is {\em normal} toric varieties. Briefly
such varieties are related to polyhedral cones and polyhedra with
integer or rational generators. The standard approach is to take the
such a cone $\sigma$ and compute its Hilbert basis, which is a set of
integer generators of the dual cone which gives all integer grid points in
that cone. From this there is a natural toric ideal. But an open
problem, it seems to the authors, is whether the rich theory of
normal and polyhedra has a role in  classical physics and
engineering.

The second area would be the natural development from the last
section. A discussion missing from  in this paper is
the way in which differentials are convert to derived quantities.
For example velocity, which is $\frac{\partial y}{ \partial t}$, for
some length variable $y$ and time $t$ is awarded the derive quantity
$LT^{-1}$. One way to keep the advantages of awarding derived
quantities to differential terms, but keep the meaning of
differentials is to use combinations of differential and polynomial
operators. The algebraic environment which combines
differential operators of this kind with polynomial algebras are differential algebras
and in particular Weyl and Orr algebras. It would be useful to develop a
type of generalization of dimensional analysis which combined
differential algebras with the invariance touched on in the last
section.

The third area is considered because the authors came to this work from the
use of experimental design methods in engineering. It seems that having easy
access to {\em all} primitive invariants should expand the scope of
experimental design methods based on invariants, which is a small but established field
see \cite{gibbings} \cite{grove}. This is mention briefly in
(\ref{wind}). The authors hope to develop this idea.

\bibliographystyle{amsplain}
\bibliography{dimension}
\end{document}